 \documentclass{aa}  

\usepackage{graphicx}
\usepackage{adjustbox}
\usepackage{amsmath}
\usepackage{color}
\usepackage{txfonts}
\usepackage[colorlinks=true,linkcolor=blue, allcolors=blue]{hyperref}
\usepackage{enumitem}
\usepackage[font=small,skip=0pt]{caption}

\begin{document}

   \title{Particle-in-cell simulations of electron--positron cyclotron maser forming pulsar radio zebras}

   \titlerunning{}

   \author{Mat\'{u}\v{s} Labaj
          \inst{1},
          Jan Ben\'{a}\v{c}ek
          \inst{2,3} and
          Marian Karlick\'{y}
          \inst{4}
          }
          
    \authorrunning{Labaj, Benáček \& Karlický}

   \institute{
   		Department of Theoretical Physics and Astrophysics, Masaryk University, 611 37 Brno, Czech Republic \\ \email{labaj@physics.muni.cz}
   		\and
            Institute for Physics and Astronomy, University of Potsdam, 14476 Potsdam, Germany
            \and 
            Center for Astronomy and Astrophysics, Technical University of Berlin, 10623 Berlin, Germany
         \and
             Astronomical Institute, Czech Academy of Sciences, 251 65 Ond\v{r}ejov,
             Czech Republic \\
             }

   \date{Received: ; accepted:}

 
  \abstract
   {The microwave radio dynamic spectra of the Crab pulsar interpulse contain fine structures represented via narrow-band quasiharmonic stripes. This pattern significantly constrains any potential emission mechanism. Similarly to the zebra patterns observed in, for example, type IV solar radio bursts or decameter and kilometer Jupiter radio emission, the double plasma resonance (DPR) effect of the cyclotron maser instability may interpret observations.
   }
   {We provide insight at kinetic microscales of the zebra structures in pulsar radio emissions originating close to or beyond the light cylinder.
   }
   {We present the first electromagnetic relativistic particle-in-cell (PIC) simulations of the electron-positron cyclotron maser for cyclotron frequency smaller than the plasma frequency. In four distinct simulation cycles, we focused on the effects of varying plasma parameters on the instability growth rate and saturation energy. The physical parameters were the ratio between the plasma and cyclotron frequency, the density ratio of the ‘hot’ loss-cone to the ‘cold’ background plasma, and the loss-cone characteristic velocity.
   }
   {In contrast to the results obtained from electron--proton plasma simulations (for example, in solar system plasmas), we found that the pulsar electron--positron maser instability does not generate distinguishable X and Z modes. On the contrary, a singular electromagnetic XZ mode is generated in all studied configurations close to or above the plasma frequency. Highest instability growth rates were obtained for the simulations with integer plasma-to-cyclotron frequency ratios. The instability is most efficient for plasma with characteristic loss-cone velocity in the range $v_\mathrm{th}=$ 0.2--0.3$c$. For low density ratios, the highest peak of the XZ mode is at the double frequency of the highest peak of the Bernstein modes, indicating that the radio emission is produced by a coalescence of two Bernstein modes with the same frequency and opposite wave numbers. Our estimate of the radiative flux generated from the simulation is up to $\sim$30\,mJy from an area of 100\,km$^2$ for an observer at 1\,kpc distance without the inclusion of relativistic beaming effects, which may account for multiple orders of magnitude.
   }
  {}


   \maketitle
%

\section{Introduction}
\label{sec:introduction}

Since their first discovery nearly 50 years ago, pulsars remain a widely studied category of astronomical objects. They possess one of the strongest magnetic fields in the known universe, enabling fascinating effects, such as the pair plasma production, which are considerably difficult to reproduce in laboratory. Although the radio emission of pulsars has been subjected to intense research \citep{sturrock:197, 1975ApJ...196...51R, 1987ApJ...320..333U, 10.1111/j.1365-2966.2009.14663.x, PhysRevLett.124.245101, Melrose2021}, there is still no general consensus on the mechanisms behind the origin of pulsar complexly structured radio emission.

Among the considered radio emission mechanisms are coherent curvature radiation \citep{1977MNRAS.179...99B,2000ApJ...544.1081M,2004ApJ...600..872G,2017JApA...38...52M}, cyclotron instability emission \citep{1991MNRAS.253..377K, 1999MNRAS.305..338L}, free electron laser \citep{2004A&A...422..817F,2021ApJ...922..166L}, relativistic plasma emission generated through relativistic streaming (beaming) instabilities \citep{1988Ap&SS.140..325U,1994ApJ...428..261W,1999ApJ...521..351M,2021A&A...649A.145M,2021ApJ...923...99B,2021ApJ...915..127B}, or linear acceleration emission,  due to  particles oscillating parallel to the magnetic field \citep{2008ApJ...683L..41B, 2009ApJ...696..320L, 2013MNRAS.429...20T,2021arXiv211105262B}. 

\subsection{Zebra pattern observed in Crab pulsar emission}
The dynamic spectra of the Crab pulsar radio emission, obtained via the radio instruments like Karl G. Jansky Very Large Array, Arecibo Observatory, and Robert C. Byrd Green Bank Telescope \citep{eilek, 2015ApJ...802..130H}, contain valuable information about the emission mechanism at plasma kinetic microscales. Remarkably, the dynamic spectra of the Crab pulsar radio interpulse are characterized by fine structures, represented via relatively narrow-band quasiharmonic stripes. The stripes are similar to Zebra patterns observed in Type~IV solar radio bursts \citep{elgaroy, 1975SoPh...44..173K, zheleznyakov:1975a} or Jovian decametric to kilometric radio emissions \citep{2016Icar..272...80L}. Specifically, the dynamic spectra from the Crab pulsar show many similarities to the solar radio bursts \citep[for example][]{2005A&A...437.1047C}. The spacing between these quasiharmonic stripes is not constant. Instead, the spacing increases with the observed frequency \citep{2003A&A...410.1011Z,eilek}. From the observations, summarized in \citet{2016JPlPh..82c6302E}, the relation between the emission frequency spacing between stripes $\Delta f_\mathrm{obs}$ and the stripe center frequency $f_\mathrm{obs}$ was found to be 
\begin{equation}
    \Delta f_\mathrm{obs} = 6\times 10^{-2} f_\mathrm{obs}.
\end{equation} 
The observed number of stripes varies, with both lower (up to ten, \citealt{eilek}) and higher (more than ten, \citealt{2010AJ....139..168H}) amounts of stripes detected. The zebra pattern in the Crab pulsar spectra is detected between 6 and 30 GHz; however, the emission becomes rare at $\sim$ 10 GHz and higher \citep{2016ApJ...833...47H}. This discovery significantly constrains any possible emission mechanism. 
\label{subsec:pre}

Beyond the light cylinder distance of the neutron star, 
the magnetic field strength reduces below $\sim\,\mathrm{10^6\, G}$, and the electron--cyclotron maser instability can be relevant. The required magnetic field for the observed frequency range $f$ = 6--10~GHz \citep{eilek} corresponds to relatively low lepton gyro-harmonics and is obtained for weaker magnetic fields B~$\sim~\mathrm{10^2}$\,G. The observed frequency $f_\mathrm{obs}$ is related to the source emission frequency $f$ by the Doppler relation
\begin{equation}
f_\mathrm{obs}=f\frac{\sqrt{1-\beta^2}}{1-\beta\cos\theta},
\end{equation}
where $\theta$ is the angle between the velocity and the direction of the emission in the observer's frame, and $\beta=V/c$ represents the emission source velocity $V$ related to the speed of light $c$. For $\cos\theta\approx 1$ and $\beta \approx 0.82$ \citep{1971Ap&SS..13...87Z}, the relation reduces to
\begin{equation}
    f_\mathrm{obs}\approx 3.2f.
\end{equation}
Assuming the emission is generated near the plasma frequency $f_\mathrm{p}$, then the plasma frequency is in the range $\mathrm{1.8-3}$~GHz \citep{2012AstL...38..589Z}, and the estimated particle number density at the source is $N \approx \mathrm{0.4-1.1\times 10^{11}\,cm^{-3}}$. The requirement of a relatively weak magnetic field and a higher particle number density could be considered as the largest caveat of the proposed model; however, the intensity of the magnetic field can significantly decrease close to the magnetospheric current sheets and at distances of several light cylinder radii where the magnetic field is partially dissipated \citep{2020A&A...642A.204C}. Proposed configurations supporting the realization of the cyclotron maser include a neutral current sheet with a transverse magnetic field or a highly elongated magnetic trap. The geometry of the source is discussed in further detail in \citet{2012AstL...38..589Z} or \citet{zheleznyakov:2020}, showing that the position of the emission bands and their frequency spacing is determined by variations in plasma density and magnetic field intensity along the current sheet, and the emission bands' fine structures depend on the plasma and magnetic field variations orthogonal to the current sheet.

\subsection{Electron-cyclotron maser and the double plasma resonance}
The electron-cyclotron maser is a process capable of producing coherent radiation from plasma. Originally proposed as a somewhat exotic concept, it gained a lot of traction during the last 30 years as a dominant mechanism for producing strong coherent emissions in collisionless magnetized plasmas \citep{1979ApJ...230..621W, 1982ApJ...259..844M}. The concept of an electron--cyclotron maser (or positron--cyclotron maser) presents a mechanism that emits radiation at frequencies near the electron cyclotron frequency and may emit at its harmonics \citep{twiss, 1959PhRvL...2..504S}. 

The maser instability is driven by a positive gradient of velocity perpendicular to the magnetic field in the velocity distribution function (VDF). The instability growth rate in the perpendicular direction to the magnetic field depends on the positive gradient of the VDF $f$ as \citep{1986ApJ...307..808W}
\begin{equation}
    \Gamma \propto \frac{1}{u_\perp}\frac{\partial f}{\partial u_\perp},
\end{equation}
where $u_\perp = p_\perp / m_\mathrm{e}$ is the perpendicular momentum per mass unit.
The maser amplification occurs due to a resonant interaction between the plasma waves and energetic, charged particles. Specifically, the resonance condition is
\begin{equation} \label{eq:res}
    \omega - k_\parallel v_\parallel - \frac{s\omega_\mathrm{c}}{\gamma}=0,
\end{equation}
where $k_\parallel$ is the parallel wave number  component of the wave vector $\vec{k} = (k_\parallel, k_\perp$) with respect to the magnetic field $\Vec{B}$, $v_\parallel$ is the parallel component of the particle velocity $\vec{v} = (v_\parallel, v_\perp)$, $\omega_\mathrm{c}$ is the particle cyclotron frequency, $s$ is the gyro-harmonic number, and $\gamma = (1 - \frac{v^2}{c^2})^{-\frac{1}{2}}$ is the particle Lorentz factor. 

Due to the double plasma resonance, the maser instability produces a sharp, exponential increase of the resonant waves in a magnetized plasma. Under the condition
\begin{equation} \label{eq:DPRcond}
    \omega_\mathrm{p}/\omega_\mathrm{c} \gg 1,
\end{equation}
when the upper hybrid frequency coincides with the harmonics of cyclotron frequency, the radiation mechanism explains the observed zebra patterns in the radiograms of the Sun, Jupiter, and particularly the Crab pulsar. The analogy between the solar radio zebra patterns and the radiation of the Crab pulsar could be interpreted via the DPR effect, suggesting that the pulsar magnetospheric plasma contains local non-relativistically hot regions with relatively higher particle number density, weaker magnetic fields to satisfy Eq.~\eqref{eq:DPRcond}, and a presence of unstable loss-cone particle distribution. Owing to the DPR effect, the electrostatic waves are enhanced at frequencies and wave numbers, which fulfill the plasma dispersion relation and the resonance condition \eqref{eq:res} in such nonequilibrium plasma. Conversion of the electrostatic waves into electromagnetic waves then leads to strong emission of radiation. According to \citet{pearlstein} and \citet{zheleznyakov:1975a,zheleznyakov:1975b,zheleznyakov:1975c}, the instability is caused by an unstable particle velocity distribution, steeply increasing  at the hybrid frequencies or wave frequencies that fulfill the resonance condition \eqref{eq:res}, such us Bernstein waves. Dispersion relations of the Bernstein modes \citep{1958PhRv..109...10B} can be obtained by solving equation
\begin{equation}
    1 - \frac{\omega_{\mathrm{p}}^2}{\omega}\frac{\mathrm{e}^{-\lambda}}{\lambda}\sum_{n=0}^{\infty}\frac{n^2I_n(\lambda)}{\omega-n\omega_\mathrm{c}}=0,
\end{equation}
where $I_n(\lambda)$ is the modified Bessel function with argument $\lambda = k_\perp^2 v_\mathrm{tb}^2/2\omega_\mathrm{c}^2$. In the limit $k_\perp \rightarrow 0,\, n=1$, and $\omega_\mathrm{p} \gg \omega_{c}$, the Bernstein dispersion branch approaches the upper hybrid branch defined as 
\begin{equation}
    \omega^2_\mathrm{UH} = \omega^2_\mathrm{p} + \omega^2_\mathrm{c} + 3k^2 v_\mathrm{tb}^2,
\end{equation}
where $v_\mathrm{tb}$ is the plasma thermal velocity.

Given that the upper hybrid frequency is close to harmonics of the electron cyclotron frequency with $s$ being the gyroharmonic number, we obtain the frequency condition from Eq.~\ref{eq:res} for instability growth as
\begin{equation}
    \omega_\mathrm{UH} \approx s \omega_\mathrm{c}.
\end{equation}
Because in the emission model, individual zebra stripes are generated at various $\omega_\mathrm{p}/\omega_\mathrm{c}$ ratios, the resonance condition can be satisfied for several sequential harmonics only in weakly magnetized plasmas where the cyclotron frequency is much lower than the plasma frequency
\begin{equation}
    \omega_\mathrm{c} \ll \omega_\mathrm{p},
\end{equation}
reducing to the approximate equality
\begin{equation} \label{eq:resonance}
    \omega_\mathrm{p} \approx s\omega_\mathrm{c}. 
\end{equation}
In this limit, the maser instability supports the growth of the electrostatic upper hybrid waves. However, a coalescence process is required to convert the electrostatic waves into electromagnetic radiation capable of escaping the source region \citep{1986ApJ...307..808W}. 

The solutions of plasma electromagnetic dispersion perpendicular to the magnetic field in electron--proton plasma are the ``fast'' X and O mode waves propagating above the X mode cutoff frequency $\omega_\mathrm{X}$ and $\omega_\mathrm{p}$ respectively, and the ‘‘slow’’ Z mode which is confined below $\omega_\mathrm{UH}$ and thus cannot leave the plasma
\begin{equation} \label{eq:uh}
    \omega_\mathrm{p} < \omega_\mathrm{UH} = \sqrt{\omega^2_\mathrm{p} + \omega^2_\mathrm{c}} < \omega_\mathrm{X}.
\end{equation}
The X and O modes are polarized perpendicular and parallel, respectively, to the plane of the wave and magnetic field vectors. Although the Z mode cannot escape the plasma, \citet{Li_Chen_Ni_Tan_Ning_Zhang_2021} found a coalescence of the Z mode with the electrostatic mode into the escaping X mode for the solar coronal case.

As the proposed interpulse emission source of pulsars is small in area, it is difficult to measure the properties of the source observationally. Numerical simulations that resolve electromagnetic and relativistic effects on  kinetic scales can grasp the system as a whole and remain self-consistent. Given the effects studied in the pulsar magnetosphere under the constraints of the electron--cyclotron maser theory \citep{zheleznyakov:2016} occur on the kinetic level, the use of the particle-in-cell method (PIC) is advantageous due to its kinetic-scale resolution. The method was successfully used to study the solar radio emission via the electron--cyclotron maser instability and double plasma resonance effects \citep{2018A&A...611A..60B, Li_Chen_Ni_Tan_Ning_Zhang_2021, ning:2021}.
However, the nonlinear evolution, mode conversion, and the resulting electromagnetic waves of the pulsar electron--positron cyclotron maser were not studied to the best of our knowledge yet.

\subsection{Aims and structure of the paper}

In this paper, we investigate the electron--positron plasma for the first time using particle-in-cell simulations that resolve the plasma kinetic microscales at which the radio emission originates.
The aims are to investigate how the instability evolves, how it differs from the electron--proton version, and what are instability radiation properties from the perspective of a potential zebra stripe radiation-producing mechanism.

In Section \ref{sec:methods}, we discuss the used particle-in-cell code, its configurations, and the motivation behind the choice of the simulated plasma parameters. We analyze the results in Section~\ref{sec:results}, discussing the impacts of varying parameters of the studied plasma (Sect.~\ref{subsec:res1}), the dispersion properties of the system and a comparison between the electron-proton and electron--positron plasma under the same conditions relevant for the studied environment (Sect.~\ref{subsec:res2}), and the evolution of the velocity distribution and its relation to the instability growth rate (Sect.~\ref{subsec:res3}). Section~\ref{sec:discussion} follows, putting our results into a broader physical context, drawing conclusions, and presenting possible future considerations.
\begin{table*}[h!]
    \centering
    \caption{Simulation parameters of each simulation cycle. (A) nine simulations covering the frequency ratio range $\omega_\mathrm{p}/\omega_\mathrm{c}$ = 10 -- 12 equidistantly with increments of $\omega_\mathrm{p}/\omega_\mathrm{c} = 0.25$; (B) four simulations with gradually increasing characteristic thermal velocity of the hot plasma component $v_\mathrm{th}$ from 0.15$\,c$ to 0.5$\,c$; (C) simulations of four different density ratios between hot loss-cone and background particles for frequency ratio values of $\omega_\mathrm{p}/\omega_\mathrm{c}$ = 3, 5, 11 (twelve simulations in total). The simulation of the electron--proton plasma used the values in column (D). }
    
    \begin{tabular}{@{} l | c  c  c  c }
     \hline \hline
    ~ & \multicolumn{1}{c}{(A)} & \multicolumn{1}{c}{(B)} & \multicolumn{1}{c}{(C)} & \multicolumn{1}{c}{(D)} \\
    \hline
    \textbf{parameter} & \multicolumn{4}{c}{\textbf{value}} \\
    \hline
    Domain size & \multicolumn{4}{c}{$6144\times 12 \times 8$}\\
    Particles/cell  & \multicolumn{4}{c}{$1100$} \\
    Time step $\Delta t$ & \multicolumn{4}{c}{$\mathrm{0.025}\,\omega_\mathrm{pe}^{-1}$} \\
    Total time $t_\mathrm{total}$ & \multicolumn{4}{c}{60 000} \\
    $n_\mathrm{th}/n_\mathrm{tb}$ & 1/10 & 1/10 & 1/1, 1/2, 1/3, 1/10 & 1/2 \\
    $v_\mathrm{th}$ & 0.2$\,c$ & $\mathrm{0.15, 0.2, 0.3, 0.5}\,c$ & 0.2$\,c$ & 0.2$\,c$ \\
    $v_\mathrm{tb}$ & \multicolumn{4}{c}{0.03$\,c$} \\
    $\omega_\mathrm{p}/\omega_\mathrm{c}$ & 10 -- 12 & 11 & 11, 5, 3 & 5\\
    Composition & e$^-$ + e$^+$ & e$^-$ + e$^+$ & e$^-$ + e$^+$ & e$^-$ + p$^+$\\
    \hline \hline
   \end{tabular}
    \label{tab:param}
\end{table*}
\section{Methods}
\label{sec:methods}
Modified TRISTAN PIC code \citep{buneman, matsumoto} is used for the simulations. The code is three-dimensional, fully electromagnetic, and includes effects of special relativity. Particle motion is resolved via the Newton--Lorentz system of equations using the Boris push algorithm \citep{1970PhRvL..25..706B}; electromagnetic fields are solved using the Maxwell equations via the Yee lattice \citep{yee}; the current deposition is done using current-conserving scheme by \citet{1992CoPhC..69..306V}. The system is evolving self-consistently. TRISTAN uses relative scales for the sake of simplifying the calculations. The computational domain is a rectangular grid with cell dimensions set to $\Delta x = \Delta y = \Delta z = \Delta = 1$. To study the given system, periodic boundary conditions are used in all three dimensions. Time discretization is $\Delta t = 1$. The size of the time step directly relates to the plasma frequency $\omega_{\mathrm{p}}$ as $\Delta t = 0.025\,\omega_\mathrm{p}^{-1}$. The value of the speed of light is set to $c=0.5$, ensuring the CFL condition holds.

We assume two species of particles, electrons and positrons, which only differ in charge sign, which is either $Q=-1$ or $Q=+1$. Initially, both particle species are loaded into the simulation in two distinct populations: a population representing the cold, background thermal plasma with the thermal (background) velocity $v_\mathrm{tb}$ and a superimposed population of the hot loss-cone plasma component with characteristic loss-cone velocity $v_\mathrm{th}$. The background particles are characterized by Maxwell-Juttner distribution. Following \citet{zheleznyakov:1975a, zheleznyakov:1975b} and \citet{1986ApJ...307..808W}, hot particles are characterized via the Dory--Guest--Harris (DGH) distribution \citep{dgh} in the form
\begin{equation} \label{eq:dgh}
    f(v_\parallel, v_\perp) \propto u^2_\perp \exp \bigg(-\frac{u^2_\parallel + u^2_\perp}{2v^2_\mathrm{th}}\bigg),
\end{equation}
where $u_\parallel = p_\parallel/m_\mathrm{e}$ and $u_\perp = p_\perp/m_\mathrm{e}$ are the longitudinal and transverse components of the particle velocity relative to the magnetic field represented in terms of relativistic momentum $\mathbf{p}=(p_\parallel, p_\perp)$ and electron mass $m_\mathrm{e}$, and $v_\mathrm{th}$ is the characteristic loss-cone velocity, at which the loss-cone distribution reaches its maximum. The DGH velocity distribution function proved to be a good approximation for particles trapped in a magnetic field mirror \citep{dgh}, characterized by a deficit of particles with small perpendicular velocities and zero mean velocity parallel and perpendicular to the magnetic field. The particles are initially distributed uniformly in space.

The simulations are carried out in four cycles, each focusing on an independent parameter or configuration of multiple variables, and how the parameter values influence the studied instability. The detailed parameter setup of the used simulations is shown in Table \ref{tab:param}.

First, the impact of the varying ratio between the plasma and the cyclotron frequency $\omega_\mathrm{p}/\omega_\mathrm{c}$ (Table~\ref{tab:param}, column A). The simulations were carried out to test whether the instability has the highest growth rate at integer values of the frequency ratio, and how increasing gyro-harmonic number $s$ affects the resulting saturation energy of the formed plasma waves. We studied the range of $\omega_\mathrm{p}/\omega_\mathrm{c}$ = 10--12, which is similar to the environment found in solar plasma \citep{Yasnov2017, Li_Chen_Ni_Tan_Ning_Zhang_2021}. The choice of the frequency ratio range is based on the assumption of reduced magnetic field intensity in the studied region in a light--cylinder distance due to magnetic reconnection and local magnetic traps \citep{2012AstL...38..589Z, zheleznyakov:2020}, although it is worth noting that the choice is somewhat arbitrary.

Second, the behavior of the instability in configurations with a varying loss-cone characteristic velocity $v_\mathrm{th}$ (Table~\ref{tab:param}, column B). This way, we investigate the behavior of the more energetic hot population of the plasma by surveying the range $v_\mathrm{th}=0.15c$--$0.5c$ and evaluate the influence of the characteristic velocity on the studied instability by assessing its growth rate and saturation value. Thermal velocity used for the background is fixed to $v_\mathrm{tb}$ = 0.03$c$, corresponding to particles with a temperature of $T \approx$ 5.3\,MK. Increasing the background plasma temperature has very little effect on the growth rate of the instability \citep{2017A&A...598A.106B, 2018A&A...611A..60B}; therefore, we do not study this effect.

Third, the impacts of the varying ratio between the  hot and background particle number density $n_\mathrm{th}/n_\mathrm{tb}$ (Table~\ref{tab:param}, column C). For the condition $n_\mathrm{th} \ll n_\mathrm{tb}$, the dispersion properties are governed by the background plasma population, while the instability is governed by the hot, nonequilibrium plasma component. However, by selecting different values of the number density ratio between the populations, we address the possibility of the hot component also influencing the dispersion properties. In our simulations, the hot component is increasingly more prevalent, from a ratio $n_\mathrm{th}/n_\mathrm{tb} = 1/10$ up to a maximum of $n_\mathrm{th}/n_\mathrm{tb}=1/1$ (total number of macroparticles remains the same). The former constraints the plasma dominated by its background component, still keeping the growth rate large enough to evolve the instability in a few thousand plasma periods.The latter can occur through magnetic reconnection as was shown, e.g., by \citet{Yao2022}. We analyze the behavior of increasing density for frequency ratios $\omega_\mathrm{p}/\omega_\mathrm{c}$ = 11, 5, 3. Similar values were also estimated in the model of the Crab zebra \citep{zheleznyakov:2020}. The intensity of the magnetic field in the simulation under the assumption of $f_\mathrm{pe}$~=~$\mathrm{1.8-3}$\,GHz \citep{2012AstL...38..589Z} is 58--97\,G for $\omega_\mathrm{p}/\omega_\mathrm{c}$ =~11; 129--214\,G for $\omega_\mathrm{p}/\omega_\mathrm{c}$ =~5; and 214--357\,G for $\omega_\mathrm{p}/\omega_\mathrm{c}$ =~3. 
    
Fourth, one referential simulation of the electron--proton plasma (Table~\ref{tab:param}, column D). Configuration of the simulation is the same as the simulation generating the strongest XZ mode from the previous cycle to compare the electron--positron and electron--proton systems, their wave regimes, and how different masses influence the dispersion properties of the studied system. The thermal velocity of the background particles listed in Table~\ref{tab:param} applies for electrons and is calculated accordingly for protons.

The choice of the studied parameters is to assess the behavior of the electron--positron cyclotron maser instability in various scenarios of physical constraints required to generate the sought radio emission. The size of the simulations is the same for all the studied cases. The used grid dimensions are 6144~$\Delta$, 12~$\Delta$, and 8~$\Delta$ in the $x$, $y$, and $z$ directions respectively. The magnetic field $\boldsymbol{B} = (0,0,B)$ has a nonzero component in the $z$ direction, as we aim to study the effects occurring in the direction orthogonal to the magnetic field \citep{zheleznyakov:2020}. Each grid cell is initiated with 1100 macroparticles at the start of the computation (summing up to the total of $\mathrm{3.2 \times 10^8}$ macroparticles), sufficient to properly distinguish physical effects from the particle noise. All simulations run for 60\,000 time steps, equal to $\omega_\mathrm{p} t = 1500$. This time is sufficient for the instability to reach saturation for all of the studied scenarios, which either enter or gradually decrease towards relaxation. 

From the simulations, the temporal evolution of the electrostatic energy $U_\mathrm{Ex}$ in the dominant direction and VDF $f(v_\perp, v_\parallel)$ were computed. For the plasma wave analysis, dispersion diagrams are obtained through the time--space Fourier transform of the electric field components.
The obtained frequency resolution is $\Delta\omega/\omega_\mathrm{p}=0.004$ and the wave number resolution is $\Delta k_\perp  \omega_\mathrm{p}/c=0.02$. 

\section{Results} 
\label{sec:results}

\subsection{Effects of varying frequency ratio, characteristic velocity, and number density ratio}
\label{subsec:res1}
\begin{figure}[t!]
    \centering
    \includegraphics[width=1\linewidth]{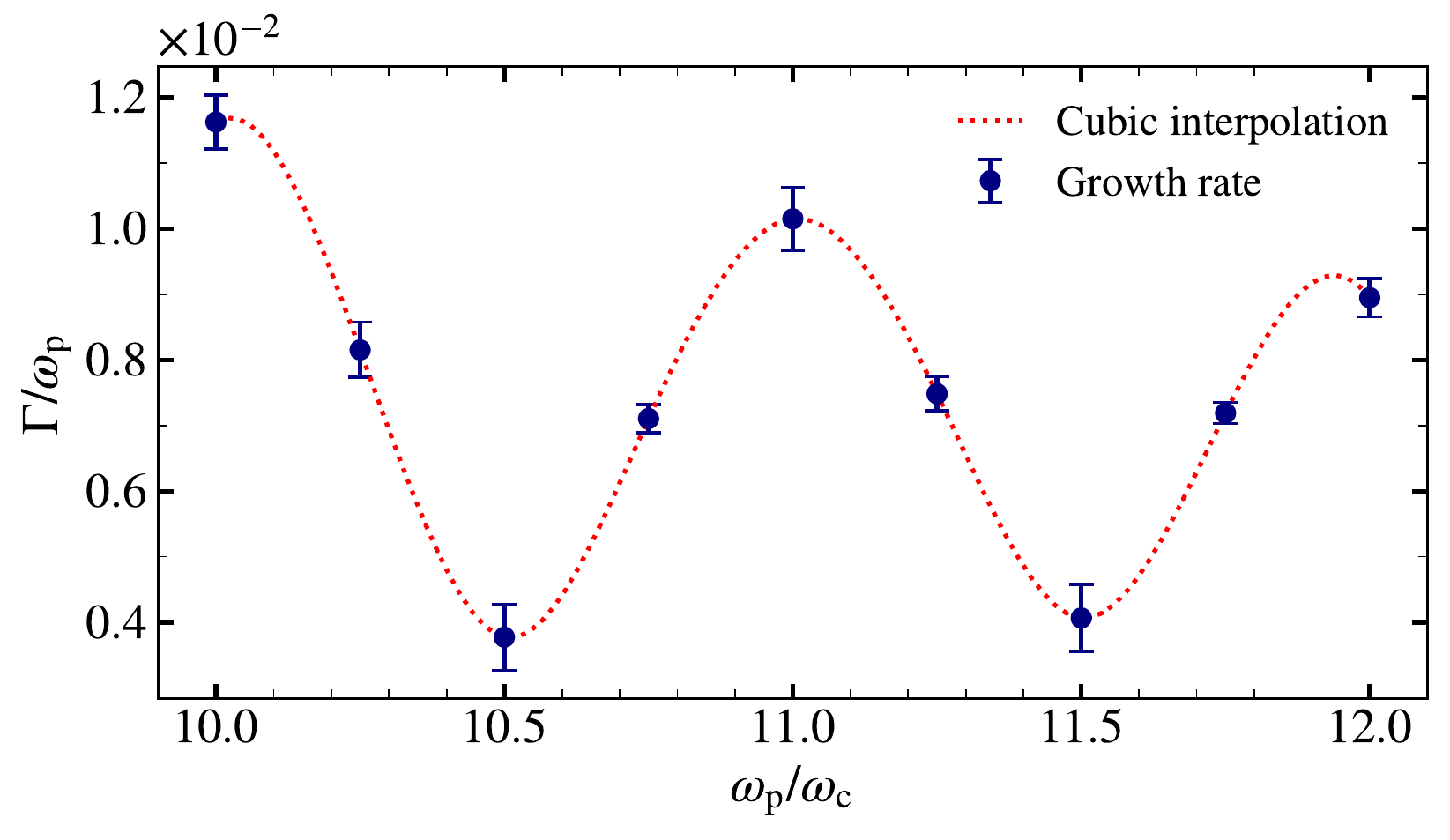}
    \caption{Dependency of the instability growth rate $\Gamma/\omega_\mathrm{p}$ on the frequency ratio $\omega_\mathrm{p}/\omega_\mathrm{c}$. Consistent with the theoretical conclusions, the maxima are located at integer values and minima between them.}
    \label{fig:growth_rate}
\end{figure}

To test the growth conditions of the electron--positron cyclotron maser with the DPR effect, we computed nine simulations (see Table \ref{tab:param}, column (A) with varying plasma-to-cyclotron frequency ratio $\omega_\mathrm{p}/\omega_\mathrm{c}$ from 10 to 12 with increments of 0.25 $\omega_\mathrm{p}/\omega_\mathrm{c}$ for each simulation. We evaluated the instability growth rate $\Gamma/\omega_\mathrm{p}$, taken as the exponential coefficient of the time--dependent electrostatic energy $U_{\mathrm{E}x}(t) \propto \mathrm{e}^{2 \Gamma t}$ computed from the $x$ component of the electric field $E$ in each grid point and integrated over the whole domain. The dependency of the growth rate $\Gamma/\omega_\mathrm{p}$ on the frequency ratio $\omega_\mathrm{p}/\omega_\mathrm{c}$ with a cubic interpolation of data in the studied interval is shown in Fig.~\ref{fig:growth_rate}. Owing to the DPR effect \eqref{eq:resonance}, the maxima are located at the integer values of the frequency ratio, and minima are in-between.

\begin{figure}[ht!]
    \centering
    \includegraphics[width=1\linewidth]{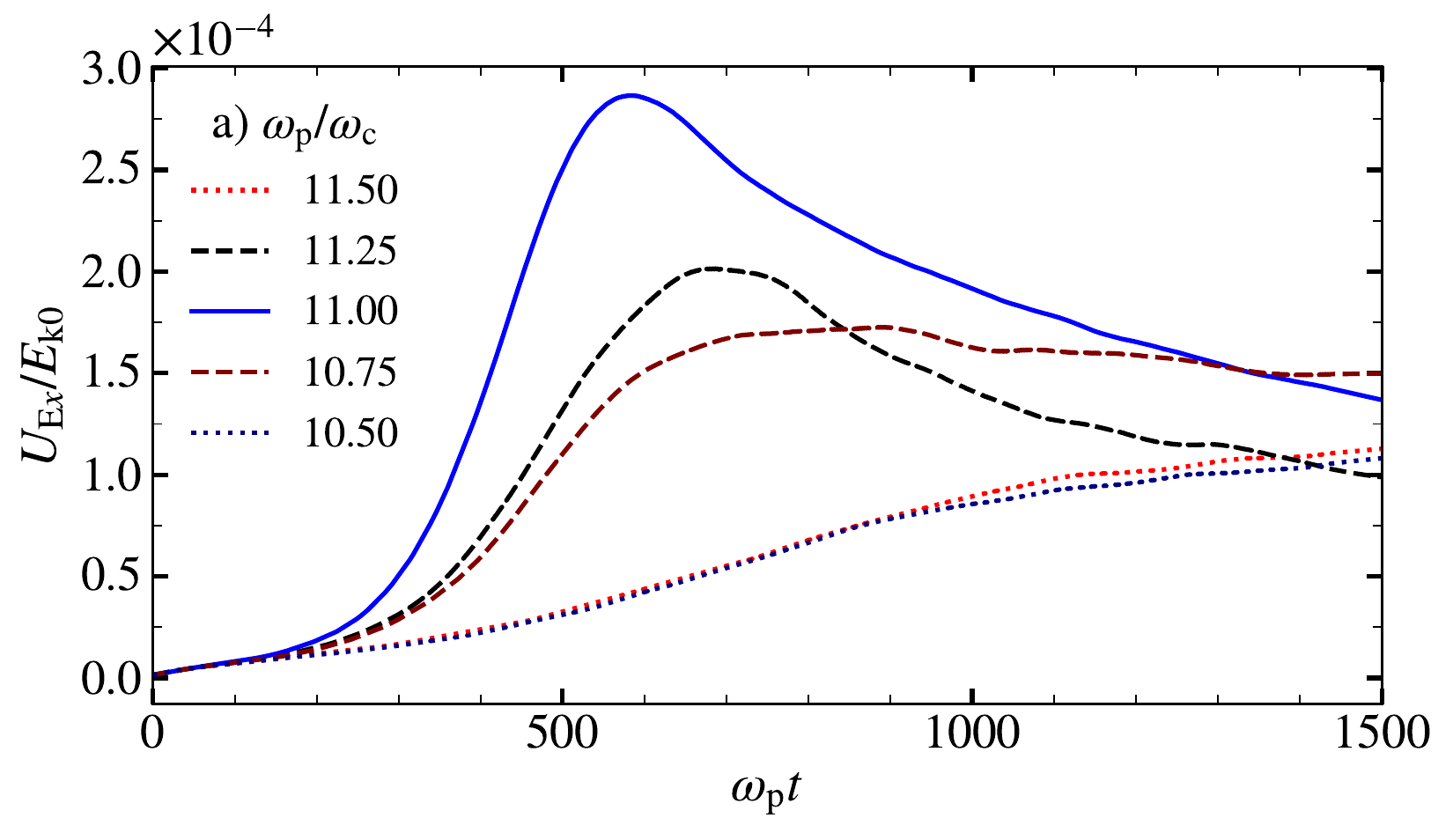}
    \centering
    \includegraphics[width=1\linewidth]{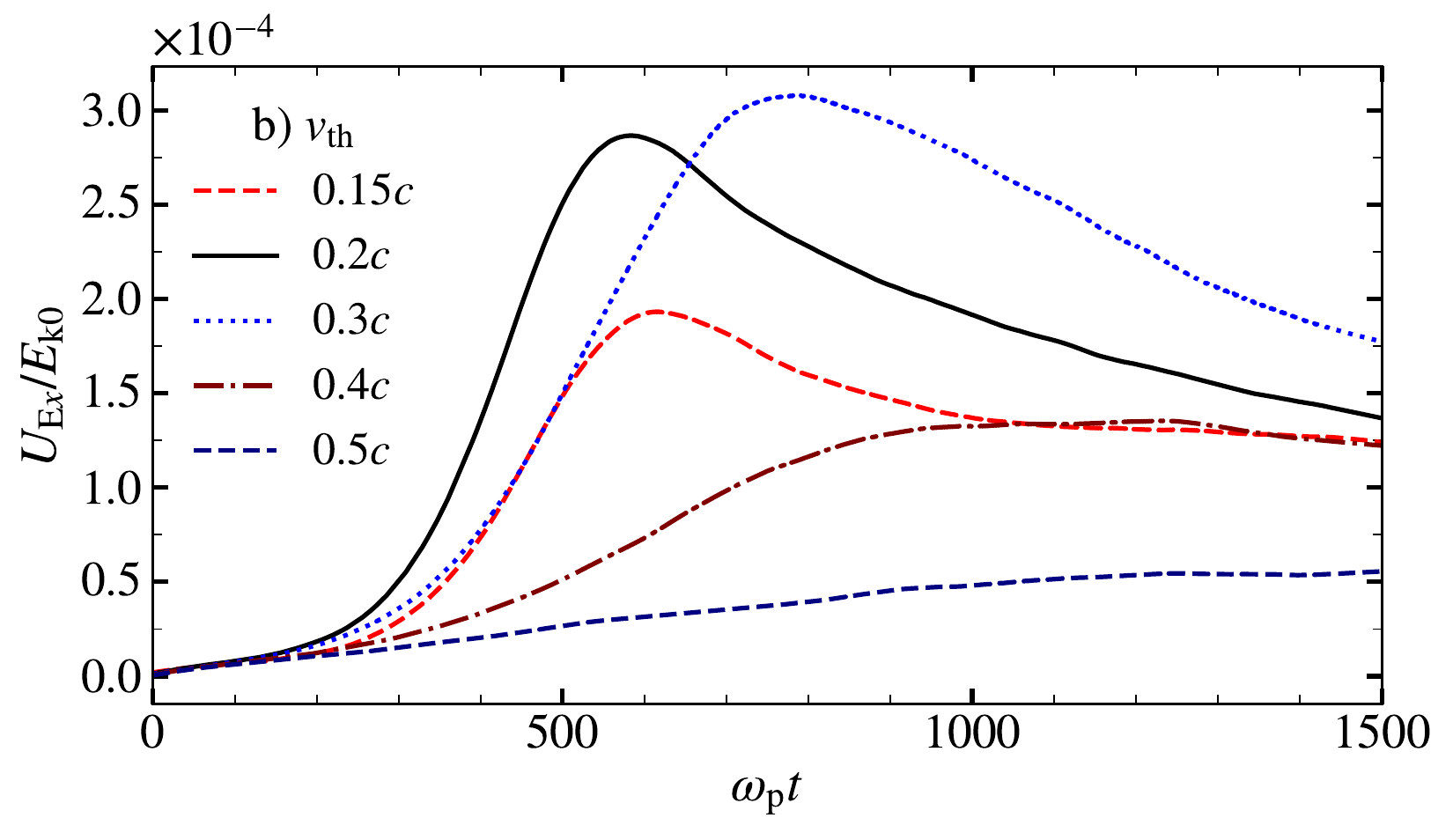}
    \centering
    \includegraphics[width=1\linewidth]{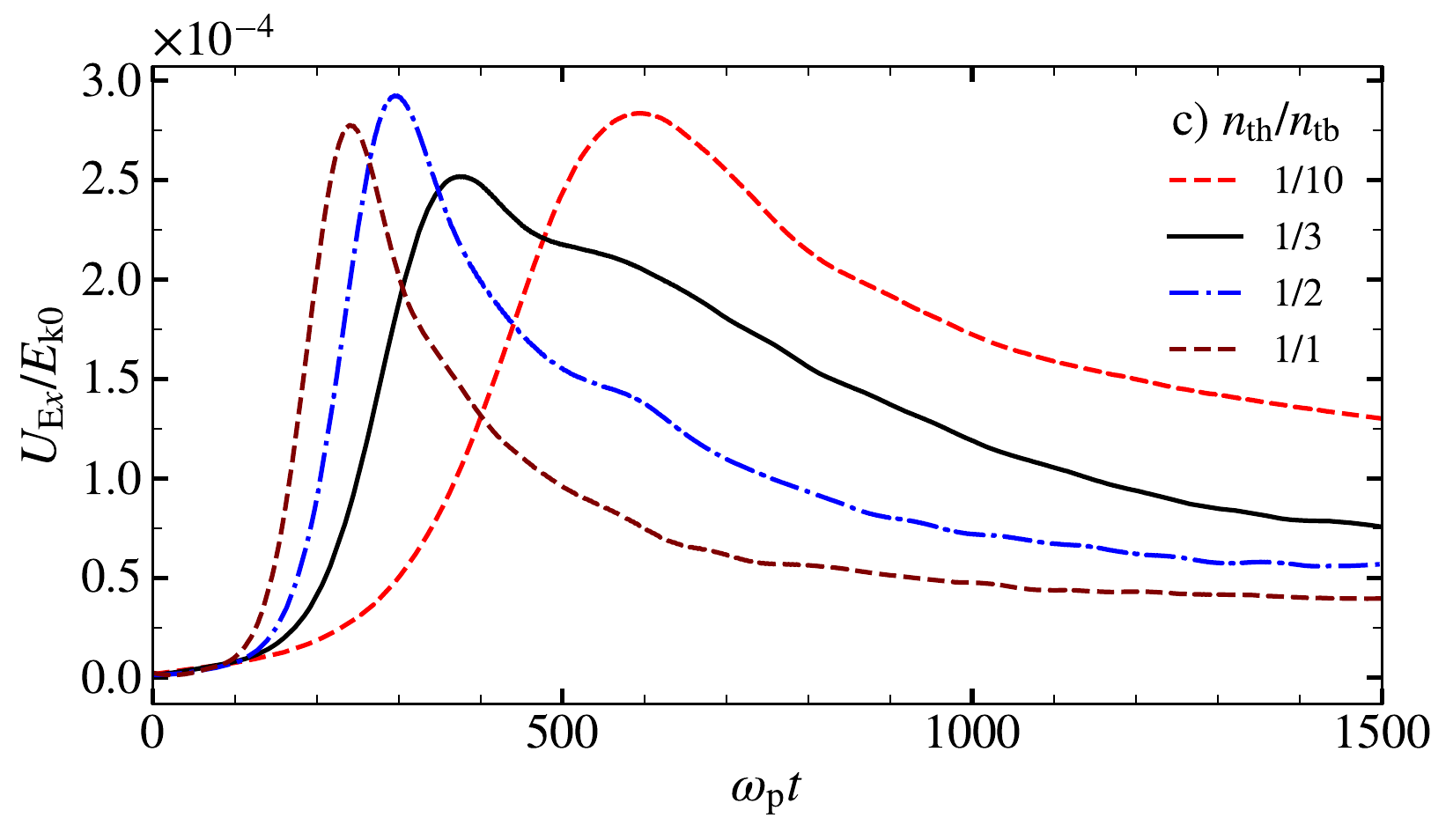}
    \caption{Evolution of the electrostatic energy of the $x$ field component $U_{\mathrm{E}x}$ scaled to the initial kinetic energy $E_\mathrm{k0}$.
    \textit{a)} Evolution of $U_{\mathrm{E}x}/E_\mathrm{k0}$, where $E_\mathrm{k0}$ is the total initial kinetic energy, for different values of the ratio between plasma and cyclotron frequency $\omega_\mathrm{p}/\omega_\mathrm{c}$. 
    \textit{b)} Evolution of $U_{\mathrm{E}x}/E_\mathrm{k0}$, for increasing value of the characteristic thermal velocity of the hot plasma component $v_\mathrm{th}$. 
    \textit{c)} Evolution of $U_{\mathrm{E}x}/E_\mathrm{k0}$ for increasing value of the ratio between the hot and background plasma density $n_\mathrm{th}/n_\mathrm{tb}$.}
    \label{fig:temp-ev}
\end{figure}

Taking a closer look at the configurations with $\omega_\mathrm{p}/\omega_\mathrm{c}$ from 10.5 to 11.5 (interval between two subsequent minima), the temporal evolution of the electrostatic energy scaled to the initial kinetic energy $U_{\mathrm{E}x}/E_{\mathrm{k0}}$ is shown in Fig.~\ref{fig:temp-ev}a. Along with the configuration with the integer value of $\omega_\mathrm{p}/\omega_\mathrm{c}$ having the highest growth rate, it also exerts the highest saturation energy.

To further build on the foundation of the configuration with the highest saturation energy ($\omega_\mathrm{p}/\omega_\mathrm{c}$ = 11, Table~\ref{tab:param}, column B), five subsequent simulations with increasing loss-cone characteristic velocity $v_\mathrm{th}$ were computed to investigate whether increased $v_\mathrm{th}$ has a positive impact on the instability growth rate and to what extent the saturation value can be increased. The temporal evolution of electrostatic energy scaled to the corresponding initial kinetic energy $U_{\mathrm{E}x}/E_{\mathrm{k0}}$ is in Fig.~\ref{fig:temp-ev}b. The configuration with $v_\mathrm{th}=0.2c$ shows the highest growth rate $\Gamma$ relative to the total kinetic energy of the simulation; however, the highest energy ratio $U_{\mathrm{E}x}/E_{\mathrm{k0}}$ is reached in configuration with $v_\mathrm{th}=0.3c$. Therefore, the highest growth rate is not necessarily accompanied by the highest saturation energy. What is apparent is that the instability grows strongly with $v_\mathrm{th}$ up to 0.3$c$. Further increasing the characteristic velocity does not increase both the instability growth rate and saturation value. The decrease of the growth rate at sufficiently high characteristic velocities implies that the efficiency of the instability is low for highly relativistic energies \citep{2011PhRvE..83e6407V}. 

Along with the study of increasing characteristic velocity $v_\mathrm{th}$, the investigation of an increasing ratio between the hot and background plasma component $n_\mathrm{th}/n_\mathrm{tb}$ for four different cases was conducted (Table~\ref{tab:param}, column C). Fig.~\ref{fig:temp-ev}c shows the temporal evolution of electrostatic energy scaled to the corresponding initial kinetic energy $U_{\mathrm{E}x}/E_{\mathrm{k0}}$ for density ratios $n_\mathrm{th}/n_\mathrm{tb}$ = 1/10, 1/3, 1/2, and 1/1. We only consider the $x$ component (with other dimension's components averaged), because we primarily focus on effects perpendicular to the magnetic field. Higher density ratios represent the possibility of a more energetic plasma environment, resulting in the instability forming faster and with a higher growth rate $\Gamma$; however, the saturation energy is comparable between all studied cases.

\begin{figure}[ht!]
    \centering
    \includegraphics[width=1\linewidth]{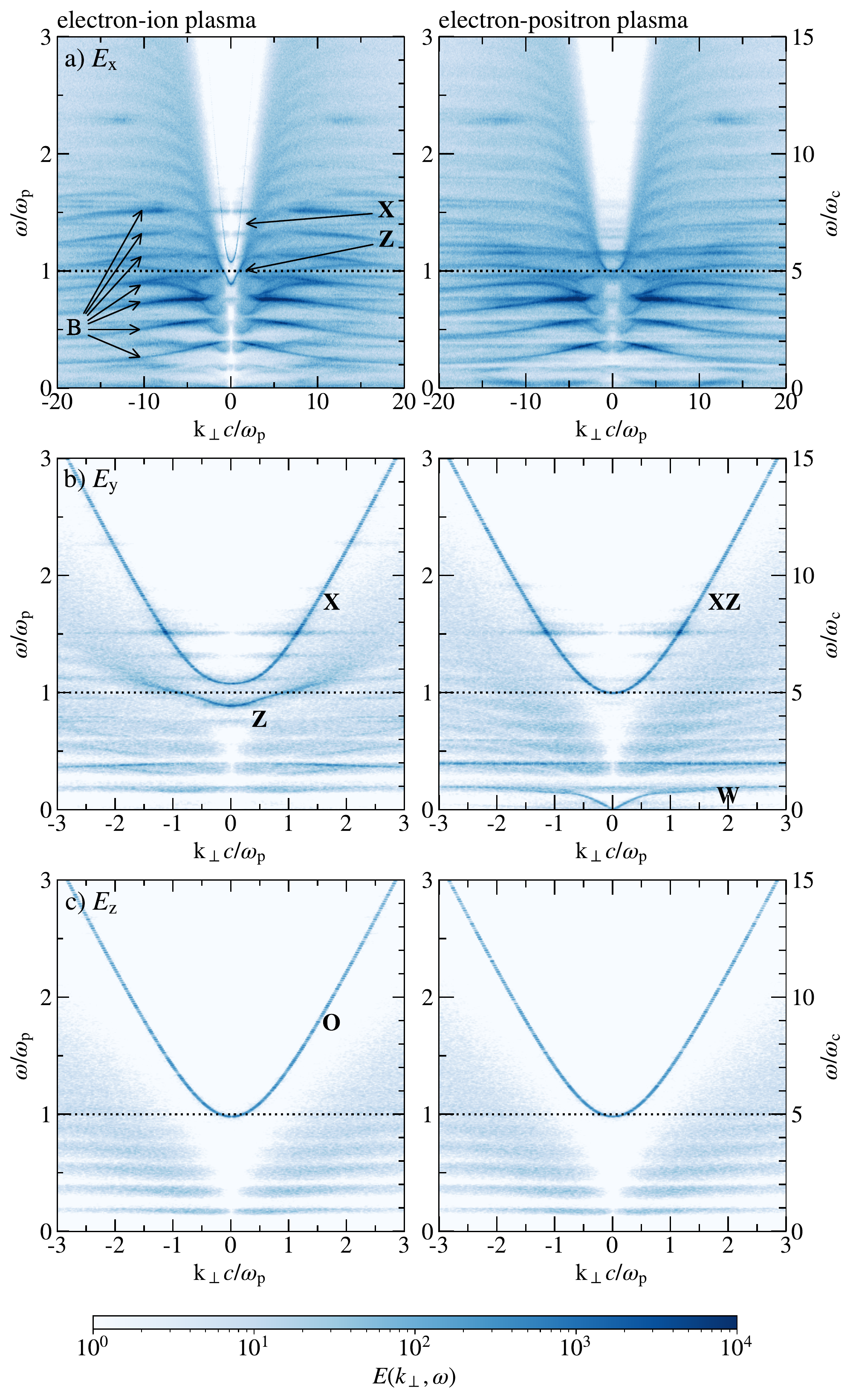}
    \caption{Dispersion diagrams for electron--proton plasma (left column) and electron--positron plasma (right column). Frequency ratio is $\omega_\mathrm{p}/\omega_\mathrm{c}$~=~5, density ratio between the hot and background plasma is $n_\mathrm{th}/n_\mathrm{tb}$ = 1/2. Rows show dispersion in the $E_x$ (longitudinal waves), $E_y$ (transverse waves), and $E_z$ components. Bernstein (B), X, Z, XZ, whistler (W), and ordinary (O) modes are denoted together with plasma frequency $\omega_\mathrm{p}$ (black dotted line).}
    \label{fig:ion}
\end{figure}

\begin{figure}[t!]
    \centering
    \includegraphics[width=0.95\linewidth]{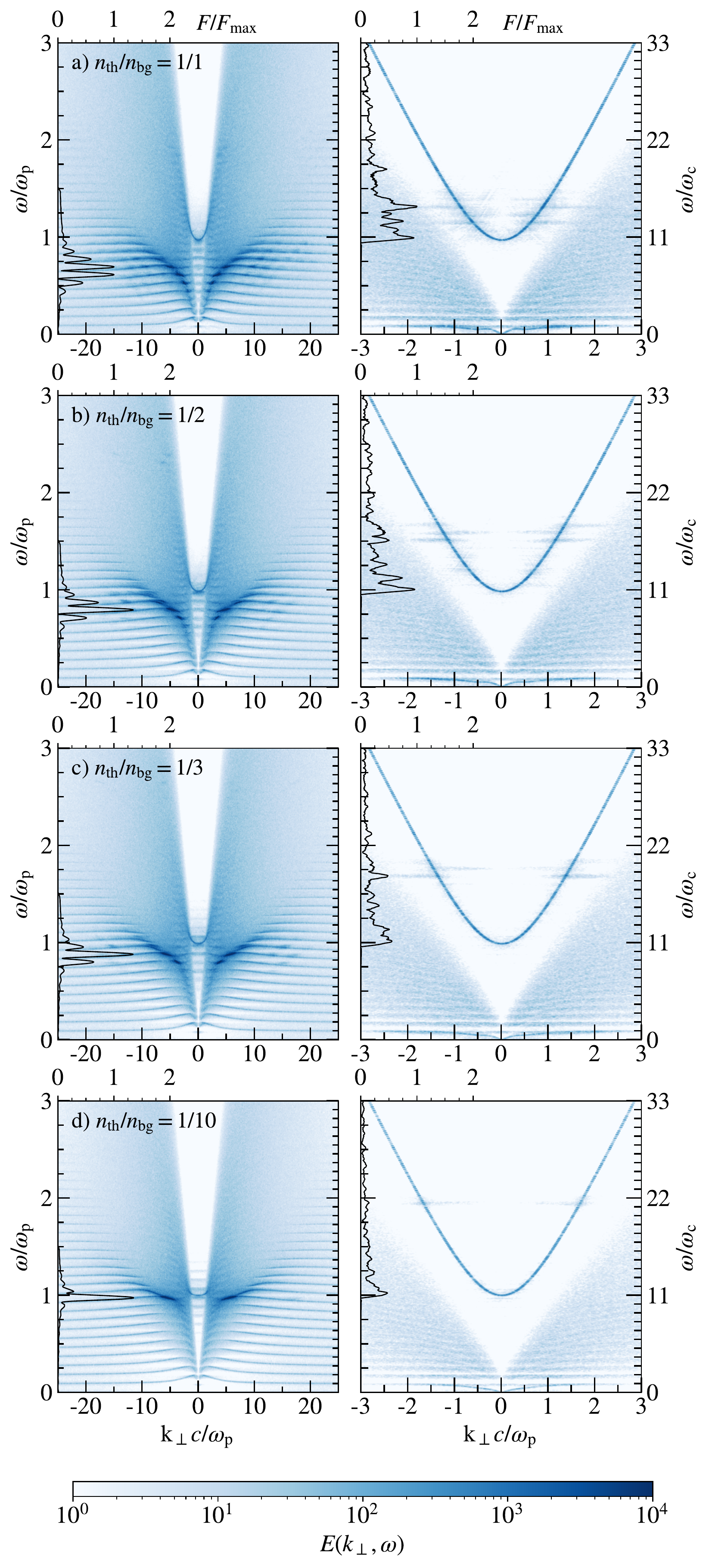}
    \caption{Dispersion diagrams for various values of $n_\mathrm{th}/n_\mathrm{tb}$ (rows) with frequency ratio $\omega_\mathrm{p}/\omega_\mathrm{c}$ = 11 for the electron--positron plasma. Left column: $E_x$. Right column: $E_y$. Dispersion diagrams are overlaid with an integrated frequency profile scaled to the maximum of the top-most case ($n_\mathrm{th}/n_\mathrm{tb}=1$).}
    \label{fig:disp11}
\end{figure}

\begin{figure}[ht!]
    \centering
    \includegraphics[width=0.95\linewidth]{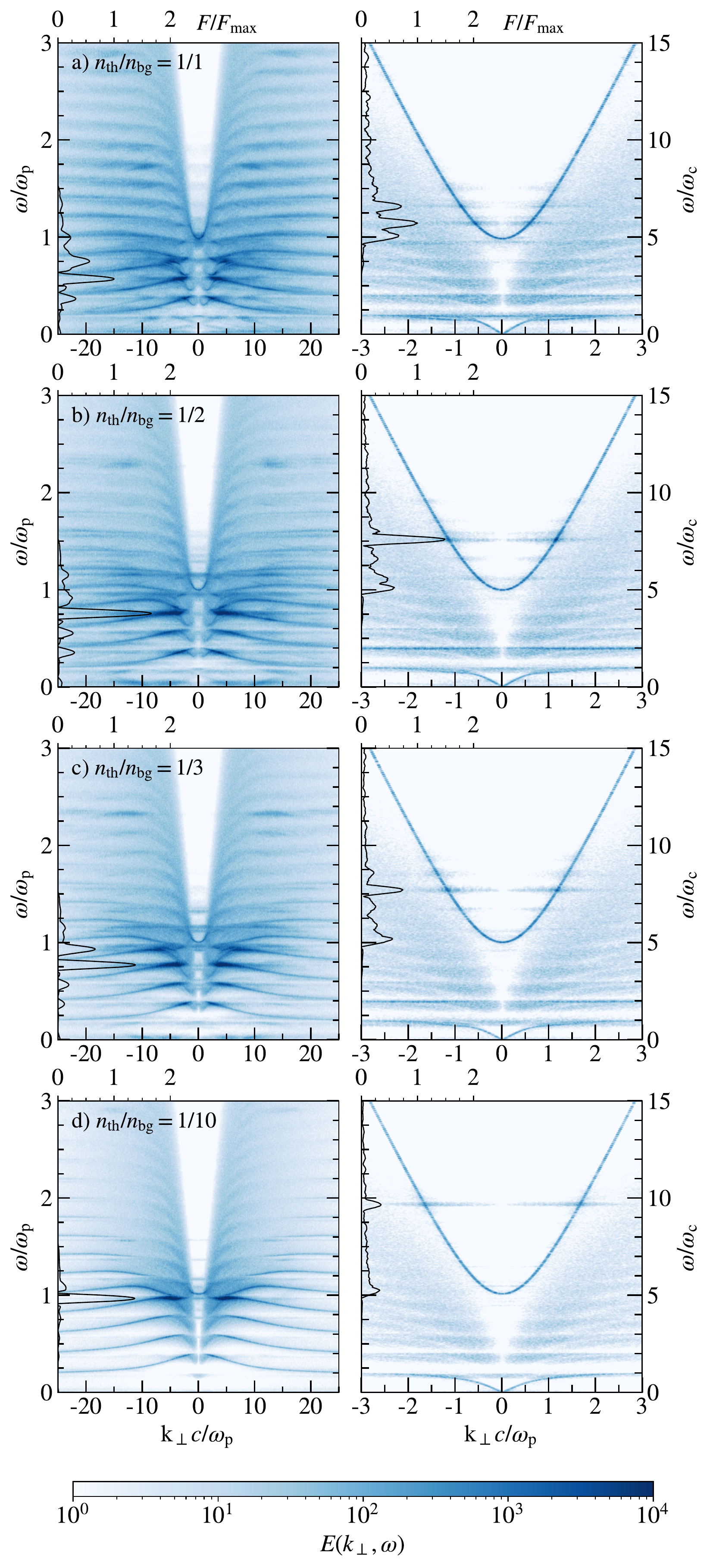}
    \caption{The same as Fig. \ref{fig:disp11}, but for $\omega_\mathrm{p}/\omega_\mathrm{c}$ = 5.}
    \label{fig:disp5}
\end{figure}

\begin{figure}[ht!]
    \centering
    \includegraphics[width=0.95\linewidth]{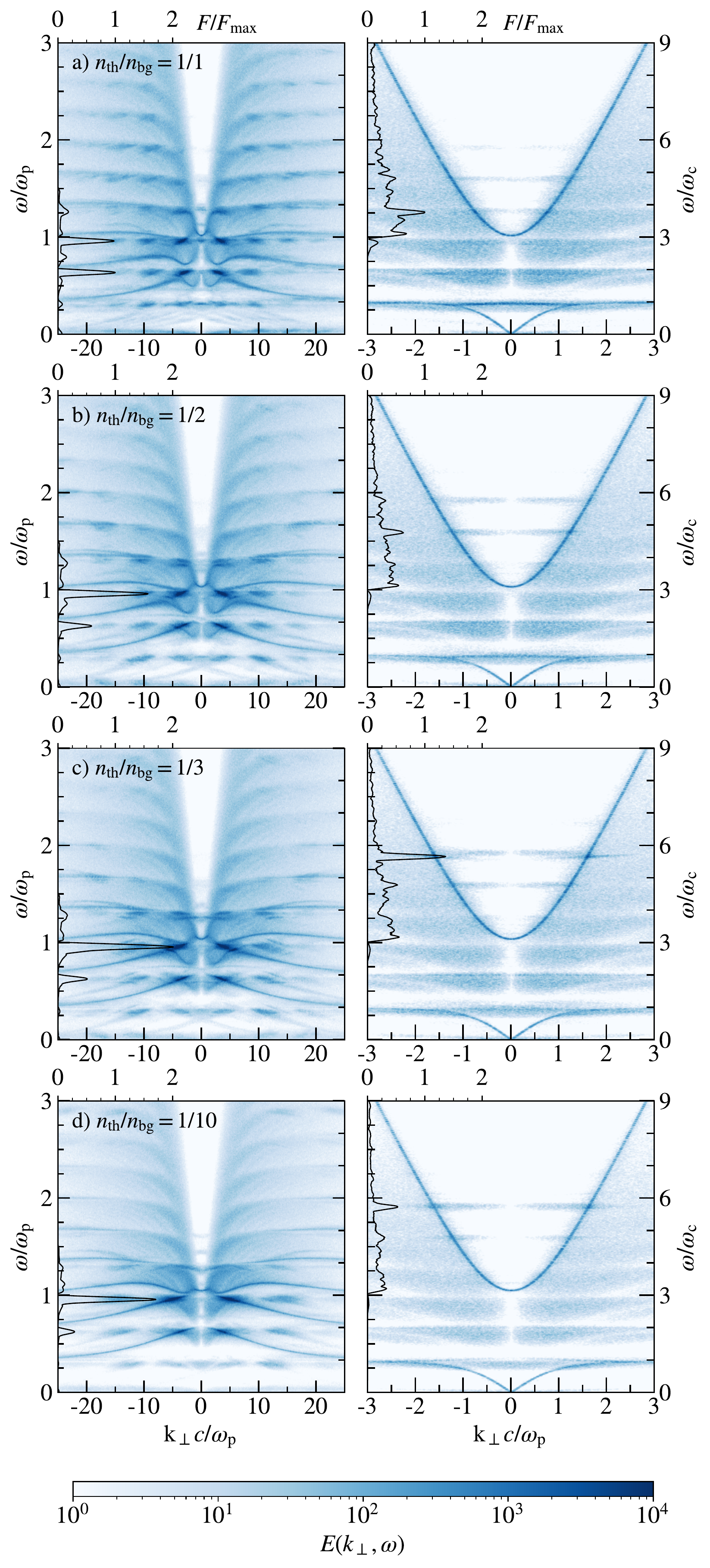}
    \caption{The same as Fig. \ref{fig:disp11}, but for $\omega_\mathrm{p}/\omega_\mathrm{c}$ = 3.}
    \label{fig:disp3}
\end{figure}

\begin{figure}[h!]
    \centering
    \includegraphics[width=1.0\linewidth]{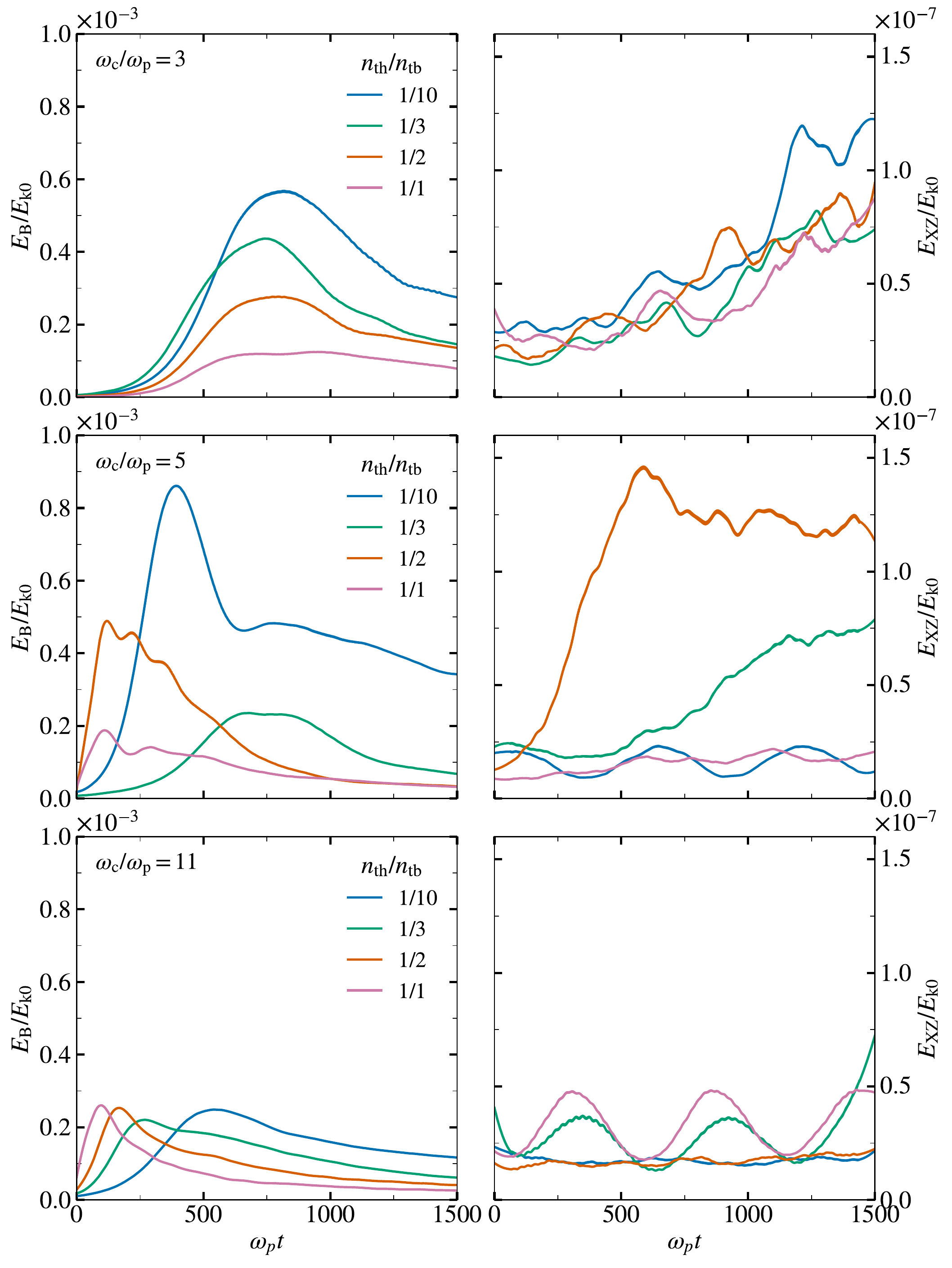}
    \caption{Temporal evolution of energy scaled to the initial kinetic energy of the electrostatic Bernstein waves (left column) and the electromagnetic XZ mode (right column) with frequency ratio $\omega_\mathrm{p}/\omega_\mathrm{c}$ = 3, 5, and 11 (rows) for different values of hot/background particle density ratio $n_\mathrm{th}/n_\mathrm{tb}$.}
    \label{fig:modes}
\end{figure}

\subsection{Wave analysis and dispersion of the plasma}
\label{subsec:res2}
The wave dispersion is obtained by analyzing the electric field $\vec{E}$ in the Fourier space. First and foremost, we compare an electron--proton plasma simulation and an electron--positron one (Table~\ref{tab:param}, column D). The configuration of these referential simulations is identical except for the mass ratio between positive and negative charge particles, where we use $m_\mathrm{+}/m_\mathrm{-}=1836$ (mass ratio between the proton and electron) for the electron--proton plasma and $m_\mathrm{+}/m_\mathrm{-}=1$ to represent the electron--positron plasma. The only parameters worth noting are the frequency ratio $\omega_\mathrm{p}/\omega_\mathrm{c}=5$ and the number density ratio $n_\mathrm{th}/n_\mathrm{tb}=1/2$. Even though the electron--positron and the electron--proton plasma have different total plasma frequencies, we assume that the contribution of the ion plasma frequency to the total plasma frequency is negligible  
\begin{equation}
    \omega_\mathrm{p} = \sqrt{\omega^2_\mathrm{pe} + \omega^2_\mathrm{pi}} = \sqrt{\omega^2_\mathrm{pe} + \frac{m_\mathrm{e}}{m_\mathrm{i}}\omega^2_\mathrm{pe}} \approx \omega_\mathrm{pe},
\end{equation}
where $m_\mathrm{e}$ and $m_\mathrm{i}$ are the electron and ion rest mass, respectively. We scale the presented quantities for both types of plasma to the total plasma frequency, as the emission occurs close to integer multiples of $\omega_\mathrm{p}$. In the case of the electron--positron plasma, scaling to $\omega_\mathrm{p}$ is justified since we do not see the separate effects of either the electron's or positron's plasma frequency in the dispersion. The comparison between the dispersion of the electron--proton and the electron--positron plasma is shown in Fig.~\ref{fig:ion}. Here, the dispersion in $x$, $y$, and $z$ components of the electric field $\mathbf{E}(k_\perp, \omega)$ is obtained through the Fourier transform of $\mathbf{E}(x,t)$. In the case of the electron--proton plasma, O, X, Z, and Bernstein wave modes are generated, yielding results similar to those obtained in solar plasma simulations \citep{Benacek2019,ni:2020,Li_Chen_Ni_Tan_Ning_Zhang_2021}. 

For our case, the most important is the dispersion in $E_y$, as it is the transverse field component in which the X and Z modes are generated. The largest difference to the electron--proton plasma is that the plasma does not generate X and Z modes separated in frequency for $k_\perp = 0$. Instead, an XZ mode is generated. This mode is generated directly at the plasma frequency $\omega_\mathrm{p}$ as opposed to the X mode being generated above and Z mode below $\omega_\mathrm{p}$ for $k_\perp = 0$ in the electron--proton plasma. Another noted difference is the presence of a whistler mode (W) in the dispersion of $E_y$ of the electron--positron plasma. Apart from these remarks, further differences between both plasma types are essentially indistinguishable, with both plasmas generating Bernstein waves in $E_x$ and O mode waves in $E_z$. The growth rate of the instability is nearly identical between both cases.

Based on the first two simulation cycles, we calculated 12 simulations across a range of different density ratios $n_\mathrm{th}/n_\mathrm{tb}$ and plasma to cyclotron frequency ratios $\omega_\mathrm{p}/\omega_\mathrm{c}$ listed in Table \ref{tab:param}(C). The dispersion in $E_\mathrm{x}$ and $E_\mathrm{y}$ components of $\vec{E}(x)$ (perpendicular to the magnetic field) are shown in Figs.~\ref{fig:disp11}--\ref{fig:disp3}. Each figure shows a dispersion with a fixed frequency ratio $\omega_\mathrm{p}/\omega_\mathrm{c}$ = 11 (Fig.~\ref{fig:disp11}), 5 (Fig.~\ref{fig:disp5}), and 3 (Fig.~\ref{fig:disp3}) for four different configurations with decreasing density ratio $n_\mathrm{th}/n_\mathrm{tb}$. The dispersion diagrams are overlaid with frequency $\omega$ integrated over wave number $k_\perp$, noted as frequency profile $F(\omega)$ of the electrostatic Bernstein waves in the longitudinal $E_\mathrm{x}$ component and the electromagnetic XZ mode in the transverse $E_\mathrm{y}$ component. 
For the integration of the wave modes, a separation of these modes from the background is required. The integration was done accurately in the case of the XZ mode, where the curve could be fitted with a hyperbole analytically and a thin region around the curve of the XZ mode with frequency half-width $0.2\,\omega_\mathrm{p}$ was selected; however, in the case of Bernstein waves there is no straightforward analytic solution for the electron--positron plasma with arbitrary density ratio; therefore, a rectangular window encompassing the maxima of the Bernstein wave intensity in the range $k_\perp c/\omega_\mathrm{p}$ = -15 to 15, $\omega$ = (0 -- 1.5)$\omega_\mathrm{p}$ was used, resulting in increased noise inclusion compared to the integration of the XZ waves. The integrated profiles $F(\omega)$ are normalized to the maximum of the top-most case $F_\mathrm{max}$, therefore $F(\omega)/F_\mathrm{max}$ is equal to 1 for the top-most case and can be higher or lower for the rest.

Lastly, we present a calculation of the estimated mode energy density for the Bernstein modes and the XZ mode. The estimated energy is obtained through the inverse Fourier transform of the mentioned modes (Bernstein modes in $E_x$ and XZ mode in $E_y$). The temporal evolution of the mode energy scaled to the initial kinetic energy of the simulations Table \ref{tab:param}(C) is shown in Fig.~\ref{fig:modes}. Most of the total electrostatic energy is carried by the Bernstein waves (noted as $E_\mathrm{B}$), as can be seen by comparing Fig.~\ref{fig:modes} (bottom--left) and Fig.~\ref{fig:temp-ev}b. There is only about a 10\% difference between the maxima of the integrated Bernstein waves and the electrostatic energy $U_{\mathrm{E}x}$ of the simulation. In other words, Bernstein waves account for 90\% of the $U_{\mathrm{E}x}$. On the contrary, the electromagnetic energy carried by the XZ modes (noted as $E_\mathrm{XZ}$) is reaching maxima that are four orders of magnitude lower than those of the Bernstein waves. However, it should be noted that the integration of the Bernstein waves was less accurate than the analytic solution of the XZ mode because the integration includes more field noise.

\begin{figure*}[h!]
    \centering
    \includegraphics[width=0.80\linewidth]{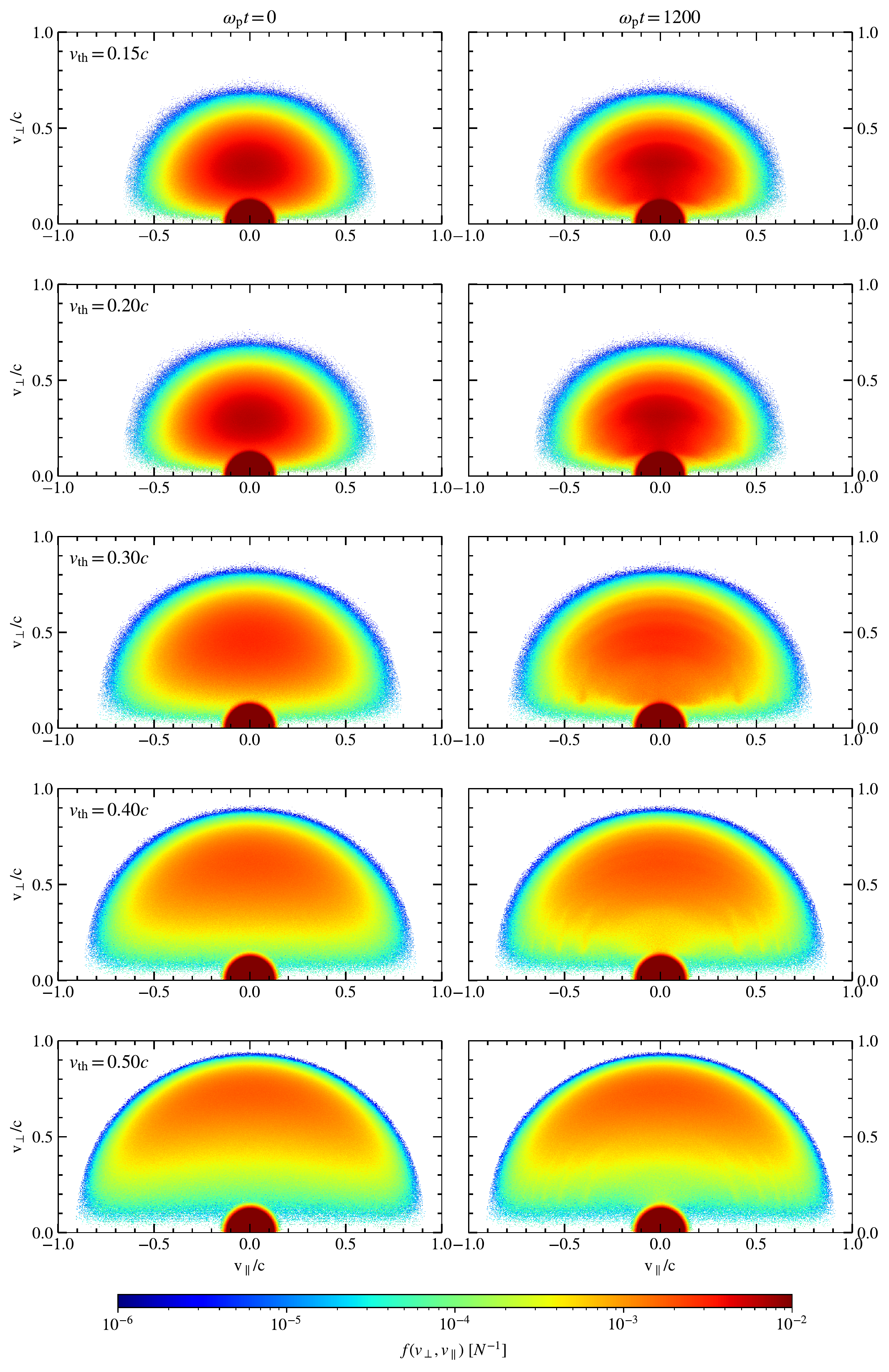}
    \caption{Velocity distribution functions at times $\omega_\mathrm{p}t$ = 0 (left column) and $\omega_\mathrm{p}t$ = 1200 (right column) for increasing value of the loss-cone characteristic velocity. The distributions are normalized to the total number of particles $N$. Combined with Fig.~\ref{fig:temp-ev}b, it is evident that the configurations with higher growth also exert more sizable changes in the velocity distribution function.}
    \label{fig:distr}
\end{figure*}

\begin{figure}[ht!]
    \centering
    \includegraphics[width=0.88\linewidth]{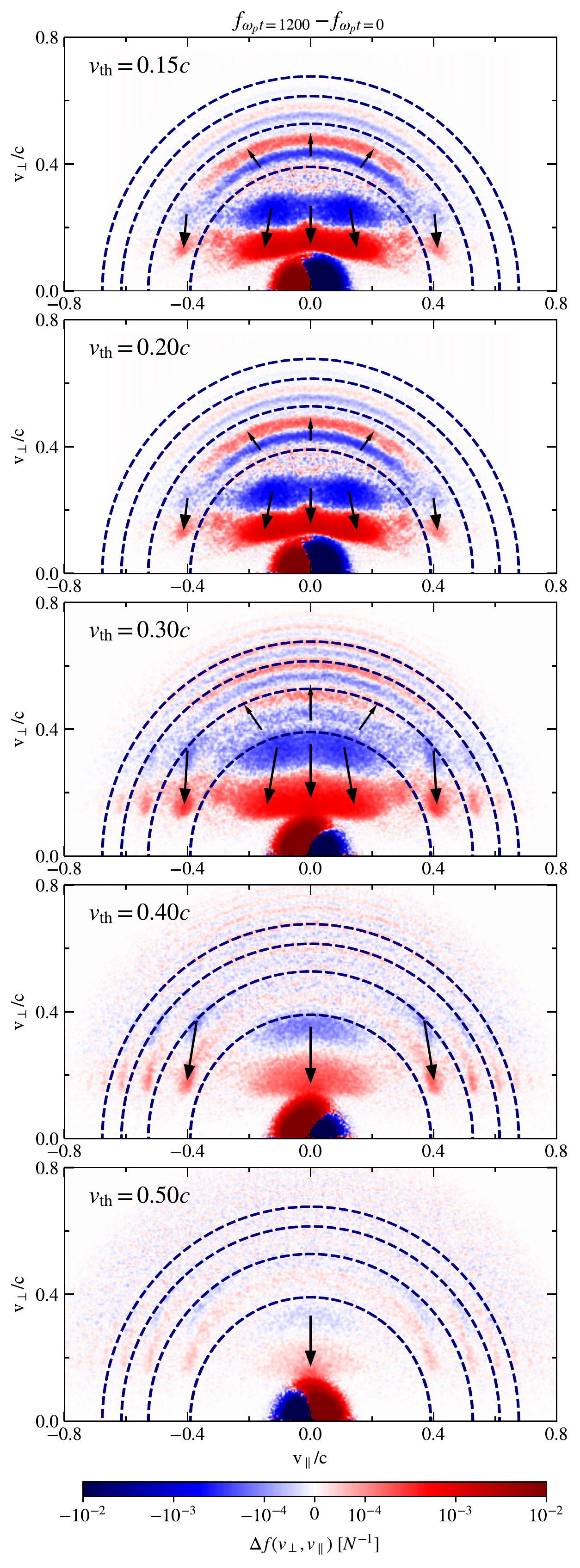}
    \caption{Differences in the velocity distribution functions $\Delta f(v_\perp, v_\parallel)$ between the final ($\omega_\mathrm{p}t=1200$) and the initial state ($\omega_\mathrm{p}t=0$) for increasing value of the characteristic thermal velocity $v_\mathrm{th}$ (rows). Dashed lines represent resonance curves for gyro-harmonic numbers $s=12,13,14,15$ with $s=12$ being closest to zero velocity. Arrows denote shifts in the VDF contained within the resonance curves.}
    \label{fig:distr_diff}
\end{figure}

\subsection{Velocity distribution function and its evolution}
\label{subsec:res3}
To comprehend and constrain the effects observed in the investigation of how the increasing characteristic thermal velocity $v_\mathrm{th}$ of the hot plasma component impacts the formation of the instability, we consider the VDF temporal evolution. The existence of a hot population described via the DGH distribution results in a positive gradient in the distribution of the perpendicular velocity $u_\perp$, producing unstable plasma via the DPR effect. The VDF $f\,(v_\perp, v_\parallel)$ (a function of perpendicular and parallel velocity component) obtained from simulations Table~\ref{tab:param}(B) (also Fig.~\ref{fig:temp-ev}b) are shown in Fig.~\ref{fig:distr}. Distributions are computed at the start of the simulation ($\omega_\mathrm{p}t=0$) and after 1200 plasma periods ($\omega_\mathrm{p}t=1200$). Combining the results from Fig.~\ref{fig:temp-ev}b and Fig.~\ref{fig:distr}, a higher growth rate is accompanied by larger shifts in the velocity distribution. For plasma with lower characteristic velocity ($v_\mathrm{th} \leq 0.3c$), the instability has a higher growth rate due to its positive gradient in $f$ being located at lower $u_\perp$. 

The evolution of the VDF is further visualized in Fig.~\ref{fig:distr_diff}, showing the difference between the initial and evolved VDF $f\,(v_\perp, v_\parallel)_{\omega_\mathrm{p}t=1200}-f\,(v_\perp, v_\parallel)_{\omega_\mathrm{p}t=0}$, overlaid with resonance curves corresponding to the solution of Eq.~\eqref{eq:resonance} for gyro-harmonic numbers $s$ = 12, 13, 14, and 15. The scenarios with lower values of $v_\mathrm{th}$ (up to 0.3$c$), for which the instability grows exponentially, also show a substantial change between the final and initial state of $f\,(v_\perp, v_\parallel)$. There is a decrease of particles delimited by the resonance curve of $s=\mathrm{12}$ towards zero perpendicular velocity; however, the particles are shifted to higher velocities periodically between the resonance curves of higher ($s$ > 12) gyro-harmonic numbers.


\section{Discussion and conclusions}
\label{sec:discussion}
The distinct zebra pattern in the Crab pulsar interpulse emission \citep{eilek} could be explained by electron--positron cyclotron maser emission driven via the double plasma resonance effect \citep{zheleznyakov:2016}. The cyclotron maser requires a relatively weak magnetic field that can be reached close to current sheets of the magnetic reconnection or in the pulsar wind beyond the light cylinder.
Using the modified 3D relativistic PIC code TRISTAN, we computed a few series of simulations, surveying the impacts of the plasma to cyclotron frequency ratio in the range $\omega_\mathrm{p}/\omega_\mathrm{c}$ = 10--12, characteristic velocity of the hot loss-cone plasma $v_\mathrm{th}$ = 0.15$c$ -- 0.5$c$, and the number density ratio between the background and hot plasma components $n_\mathrm{th}/n_\mathrm{tb}$ = 1/1 -- 1/10 for frequency ratios $\omega_\mathrm{p}/\omega_\mathrm{c}$ = 3, 5, and 11.

Comparing the simulation of the electron--proton plasma with the electron--positron plasma (Fig.~\ref{fig:ion}) for similar plasma parameters, distinguishable X (cutoff above $\omega_\mathrm{p}$) and Z (cutoff below $\omega_\mathrm{p}$) modes are generated in the electron--proton plasma. In the case of the electron--positron plasma, only one XZ mode is generated. This XZ mode approaches $\omega_\mathrm{p}$ for $k_\perp = 0$ with a slight dependence of the cutoff frequency on the density ratio $n_\mathrm{th}/n_\mathrm{tb}$ between the hot and background plasma components. Similar dependency of the decreasing cutoff frequency wave mode with increasing electron temperature is also found in the analytic study by \citet{2019JPlPh..85c9005R} for relativistically hot plasmas. Although the X mode moves to higher frequencies and Z mode to lower frequencies depending on the gyrofrequency in electron--proton plasmas \citep{Li_Chen_Ni_Tan_Ning_Zhang_2021}, the XZ mode is generated exclusively near the plasma frequency of the electron--positron plasma. 

We found that the growth rate of the instability is the highest at integer values of plasma to cyclotron frequency ratio. Our further evaluation of the instability shows that the highest saturation energy, normalized to initial kinetic energy, is reached for loss-cone characteristic velocities $v_\mathrm{th}$ in the range 0.2--0.3$c$. The highest saturation energy was obtained in simulation with $v_\mathrm{th}=0.3c$ while the highest growth rate was obtained for $v_\mathrm{th}=0.2c$. This discrepancy suggests that the highest saturation of the electrostatic energy is not necessarily accompanied by the highest instability growth. Furthermore, increasing the characteristic velocity of the loss-cone component diminishes the instability growth, showing the mechanism is less efficient for relativistically hot plasma where higher energy densities are obtained than for lower characteristic velocities. 

We observe a correlation between the growth rate of the instability and the evolution of the VDF $f\,(v_\perp, v_\parallel)$. The highest growth rates are accompanied by the most changing distributions over the course of the simulation. While some particles shift to lower perpendicualr velocities, others shift to higher velocities. These changes in the VDF are encompassed by DPR resonance curves. 

Even though the instability growth rate peaks are found at integer values of the frequency ratio $\omega_\mathrm{p}/\omega_\mathrm{c}$, this does not necessarily imply that the strongest intensity of the electrostatic and electromagnetic waves is located at frequencies of the cyclotron harmonics. The intensity maxima of the electromagnetic waves, determined in the dispersion of the electric field components (Fig.~\ref{fig:disp11}, \ref{fig:disp5}, and \ref{fig:disp3}), are located mainly at frequencies between the gyro-harmonics. Furthermore, these maxima of wave intensity in the $\omega$--$k_\perp$ domain are contained within a frequency range $\omega$ = 1--2 $\omega_\mathrm{p}$.

The particle number density ratio shifts in the phase space (see Fig.~\ref{fig:distr_diff}) with the resonance curves of the gyro-harmonics slightly above the plasma frequency, and their intensity gradually decreases with increasing the gyro-harmonic number $s$, effectively constraining the emission within the resulting frequency range. Thus, the electron--positron maser instability driven via the DPR effect could be the mechanism responsible for the generation of narrowband emission observed in the dynamic spectra of the Crab pulsar \citep{eilek}. 

\subsection{Constraints on radio emission region}
With the currently considered model of the zebra emission, we can constrain the density ratio in the emission region.
Because the frequency distances between observed zebra stripes are not constant, the individual stripes are generated in various plasma regions as the ratio between the plasma density and cyclotron frequency changes.
That implies that each emission region emits at only one specific frequency, and the observed zebra pattern is produced as a compilation of several emission regions.
Therefore, the maser mechanism also has to emit at one specific frequency.
Assuming the XZ mode emission, the density ratio $n_\mathrm{th}/n_\mathrm{tb} \lesssim 1/10$ is required to fulfill the condition.
On the other hand, if a zebra pattern with an equidistant frequency distance between stripes is observed, the emission can be produced by one plasma region with a high density ratio $n_\mathrm{th}/n_\mathrm{tb} \gtrsim 1/1$.
Moreover, the density ratio increases with the number of detected stripes.

The emission generation is a two step process. First, the kinetic instability excites arbitrary electrostatic wave modes. Then, the electromagnetic waves are generated through a coalescence of the electrostatic modes. 
In the solar plasma, a coalescence of Z mode and Bernstein (UH) mode accounts for the cyclotron maser emission  \citep{ni:2020}.
In the electron--positron plasma, a coalescence of two Bernstein modes \citep{2005A&A...438..341K} into the XZ mode may account for the observed quasi-harmonic emission of the pulsar radio interpulse. This can be seen in Fig.~\ref{fig:disp5} b, where the highest peak of the XZ mode is at the double frequency of the highest peak of the Bernstein modes. It indicates that the radio emission is produced by a coalescence of two Bernstein modes with the same frequency and opposite wave numbers. The coalescence process, however, is ineffective, as the XZ mode reaches energies approximately four orders of magnitude lower than those of Bernstein modes (Fig.~\ref{fig:modes}). Moreover, since our PIC simulation is grid based, any coalescence based process is inherently less effective due to a limited set of wave vectors $\Vec{k}$ included in the simulation. Because the emission region generates the emission at one frequency, other zebra stripes need to be generated at different
locations, which explains the non-constant spacing between zebra stripes.

\subsection{Estimation of flux density} 
To present an estimate of the generated flux density of the electromagnetic emission, we base our calculation on the simulation with the strongest XZ mode ($\omega_\mathrm{p}/\omega_\mathrm{c}~=~5$, $n_\mathrm{th}/n_\mathrm{tb}~=~1/2$). We estimate the radiative flux under the following assumptions: 1) scaling the energy of the XZ mode to the total kinetic energy of the simulation, 2) an area of the source as 100 km$^2$, 3) the distance to the source is 1\,kpc (distance of the Crab pulsar is approximately 2\,kpc), and 4) neglecting the radiative transfer effects. The kinetic energy of a particle is given by 
\begin{equation}
    E_\mathrm{k} = (1-\gamma)m_\mathrm{e}c^2.
\end{equation}
Considering the upper limit of the particle number density \citep{zheleznyakov:2016} $n = 1.2 \times 10^{11}$ cm$^{-3}$, the estimated total kinetic energy density is 
\begin{equation}
    \rho(E_\mathrm{k}) = \frac{1}{3}\rho(E_\mathrm{k, th}) + \frac{2}{3}\rho(E_\mathrm{k, bg}) = 705\, \mathrm{erg}\cdot\mathrm{cm^{-3}}.
\end{equation}
The highest energy of the XZ mode obtained from the simulations is $\approx$ $1.5\times 10^{-7}\, E_\mathrm{k}$. Under the above-mentioned assumptions, we obtain a radiative flux $P \sim 1$ mJy. Taking into account the emission source velocity ($\beta = 0.82$), the received flux is $P \sim 30$~mJy. This value is considerably lower than the observed radiative flux density \citep{eilek} by approximately 4 to 5 orders of magnitude; however, one should bear in mind that the domain size of the presented simulations was small ($\approx 350\, \mathrm{cm^3}$) and effectively 1D. On the other hand, 2D or even 3D simulations could simulate the effects of an arbitrary propagation angle instead of limiting us to the perpendicular propagation to the magnetic field. Furthermore, due to the grid-based nature of our simulation, any coalescence process is less effective. Relatively low particle number density can be responsible for the widening of the frequency peaks due to the effects of the numerical noise. 

The Doppler effect could increase the power of the emitted waves if the emission source moves toward the observer with a higher velocity than assumed above. The Lorentz factor for such a Doppler boost can be obtained in the order of $\gamma \sim 10^2 - 10^4$ in the pulsar wind \citep{Lin2023}. That could lead to an increase of the emitted power by factor of $\propto \gamma^3$ for $\gamma \gg 1$ \citep[Chapter 4.8]{Rybicki2004}. This way, for $\gamma \sim 10^3$ we can obtain the emission power of $\sim$10$^{6}$\,Jy. However, we must note that the Doppler effect does not only increase the power, but it also increases the wave frequency. Hence, for the same observed frequency, the Doppler frequency shift would require wave emission at frequencies much lower than the plasma frequency, much lower plasma frequencies, or kinetic and `XZ' mode energy densities higher than estimated for our case.

We conclude that the detected emission power from our PIC simulations is too low to explain the observed intensities of the zebra pattern.
To obtain the observed intensities, the area of the emission region or the conversion efficiency from particles to the electromagnetic waves is required to be $\sim 5$ orders of magnitude higher.
Therefore, future considerations of the electron--cyclotron maser instability should include two--dimensional analysis of the ($k_\perp$, $k_\parallel$) space to determine whether the radiation is exclusively generated in a wide or narrow angles, how the radiative power depends on the emission angle, and what the contribution by the relativistic beaming effect is. Furthermore, larger domain size would allow much less limited set of $\Vec{k}$ vectors, which might increase the efficiency of the coalescence process.

\begin{acknowledgements}
M. L. acknowledges the GACR-LA grant no. GF22-04053L for financial support.
J. B. acknowledges support from the German Science Foundation (DFG) projects BU~777-17-1 and BE~7886/2-1. M. K. acknowledges support from the GACR grant no. 21-16508J. This work was supported by the Ministry of Education, Youth and Sports of the Czech Republic through the e-INFRA CZ (ID:90140).
The authors gratefully acknowledge the Gauss Centre for Supercomputing e.V. (\href{www.gauss-centre.eu}{www.gauss-centre.eu}) for funding this project by providing computing time on the GCS Supercomputer SuperMUC at Leibniz Supercomputing Centre (\href{www.lrz.de}{www.lrz.de}).

\end{acknowledgements}

%
%

\bibliographystyle{aa}
\bibliography{references}

\end{document}